# 3-DUSSS: 3-Dimensional Ultrasonic Self Supervised Segmentation

Shaun McKnight[1*], Vedran Tunukovic[1], Amine Hifi[1], S. Gareth Pierce[1], Ehsan Mohseni[1], Charles N. MacLeod[1], Tom O'Hare[2]

[1]*Sensor Enabled Automation, Robotics, and Control Hub (SEARCH), Centre for Ultrasonic Engineering (CUE), Electronic and Electrical Engineering Department, University of Strathclyde, Glasgow, UK*

[2]*Spirit AeroSystems, Belfast, UK*

## Abstract

This study introduces a novel self-supervised learning approach for volumetric segmentation of defect indications captured by phased array ultrasonic testing data from Carbon Fiber Reinforced Polymers (CFRPs). By employing this self-supervised method, defect segmentation is achieved automatically without the need for labelled training data or examples of defects. The approach has been tested using artificially induced defects, including back-drilled holes and Polytetrafluoroethylene (PTFE) inserts, to mimic different defect responses. Additionally, it has been evaluated on stepped geometries with varying thickness, demonstrating impressive generalization across various test scenarios. Minimal preprocessing requirements are needed, with no removal of geometric features or Time-Compensated Gain (TCG) necessary for applying the methodology. The model's performance was evaluated for defect detection, in-plane and through-thickness localisation, as well as defect sizing. All defects were consistently detected with thresholding and different processing steps able to remove false positive indications for a 100% detection accuracy. Defect sizing aligns with the industrial standard 6 dB amplitude drop method, with a Mean Absolute Error (MAE) of 1.41 mm. In-plane and through-thickness localisation yielded comparable results, with MAEs of 0.37 and 0.26 mm, respectively. Visualisations are provided to illustrate how this approach can be utilised to generate digital twins of components.

## 1. Introduction

Composite materials, particularly Carbon Fibre Reinforced Polymers (CFRP), are widely utilised in industries such as aerospace, automotive, marine, and civil engineering. In modern commercial aircraft like the Airbus A350 and Boeing 787, CFRP can account for over 50% of the final structure's weight. In smaller aircraft like private jets and helicopters, CFRP usage can even reach 70-80% by weight [1], [2]. The manufacturing process involves layering multiple carbon ply sheets, preforming, and curing using a thermoset polymer in a mould. The inherent anisotropy of CFRP, resulting from the direction of fibre filaments in the weaving patterns and layup sequences, allows for precise engineering to meet specific structural requirements. This makes CFRP ideal for high-performance applications with significantly reduced weight [3], [4], [5], [6], [7], [8], [9], [10], [11]. However, the complex manufacturing process of CFRP components can introduce defects that compromise structural integrity and performance [3], [4], [6], [9], [10], [12], [13]. These defects range from delamination's and cracks to foreign object inclusions, ply stacking errors, fibre distortions, and porosities [8], [13]. Given the increasing use of composites in safety-critical components, identifying, characterizing, and quantifying defects is crucial for ensuring structural integrity and performance [6].

Non-Destructive Evaluation (NDE) comprises various methodologies utilized for examining components without compromising their integrity. Among the most prominent NDE techniques are radiography, thermography, electromagnetic approaches, and ultrasound. The choice of the most suitable NDE technique depends on the inherent characteristics of the component under examination, logistical requirements of the inspection, and the specific defects targeted for detection.

Ultrasonic Testing (UT) has become widely adopted and standardised for bulk inspections of composite components in the aerospace industry. UT's capabilities extend to detecting a wide range of volumetric defects [4], [8], [11], [12], while being relatively straightforward to implement and free from hazards compared to radiography. UT operates on the principle of transmitting, propagating, and receiving ultrasonic waves. It is commonly used for through-thickness inspection by exciting a sound wave on the surface of a component. These waves propagate through the material, and the resulting internal reflections and scatterings provide valuable insights into the component's volumetric integrity.

Ultrasonic phased arrays are often preferred for the transmission and reception of acoustic waves due to their operational flexibility, beamforming capabilities, larger coverage area, and reduced inspection times. These arrays consist of independently controllable ultrasonic transducers, enabling various electronic scanning and imaging capabilities. These capabilities include beam steering, dynamic depth focusing, and variable sub-apertures [10]. With control over each individual element or sub-aperture within a linear phased array, depth-wise sectional images, known as B-scans, can be generated from a single array (Figure 1). When combined with mechanized scanning conducted perpendicular to the linear phased array's length, three-dimensional (3D) volumetric scan data can be produced by stacking multiple individual B-scans at known positions. Volumetric ultrasonic data is commonly presented in the form of plan view (C-Scan) or sectional (B-scan) images [14].

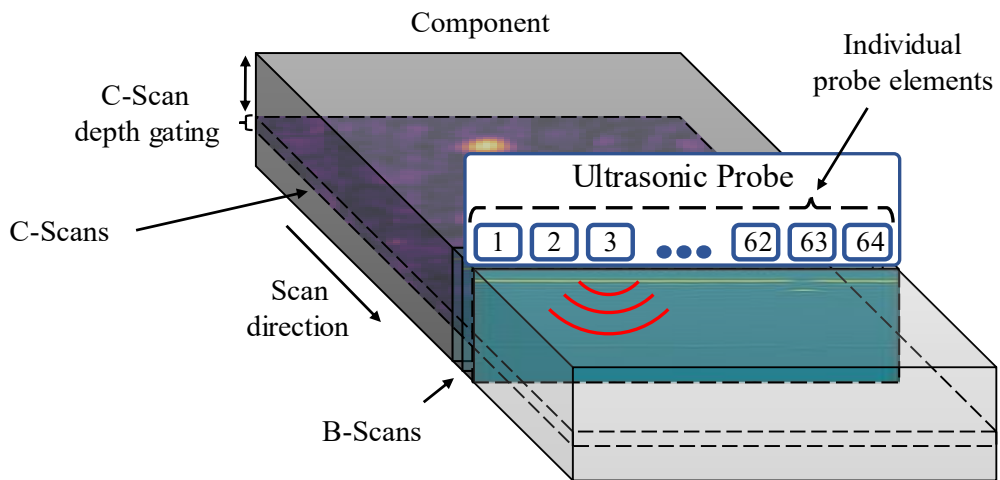

*Figure 1: Demonstration of how individual probe elements comprise a linear phased array for production of B-scan and C-scan images.*

The integration of robotics into NDE has revolutionised the inspection of large-scale components by significantly increasing automation [15]. Robotic scanning offers several advantages over manual scanning, including enhanced positional accuracy, repeatability, and a substantial reduction in scan time [16]. However, despite the advancements in robotic scanning, the analysis of results in industrial settings remains a laborious and time-intensive task. This necessitates the presence of highly trained and certified NDE operators to ensure that results meet established standards [11], [17], [18], [19], [20], [21]. Human interpretation poses two significant drawbacks: poor time efficiency and the risk of human error [19]. Therefore, there is a pressing need for research and development of automated data interpretation methods that can complement human interpretation. By reducing reliance on human interpretation, automation in NDE enhances efficiency, reliability, consistency, repeatability, and traceability, while simultaneously reducing inspection durations and costs [21], [22].

While defect detection is essential in NDE processes, it's equally important to emphasize the accuracy of localization and sizing. Traditionally amplitude drop methods, such as 6 dB drop, are widely and generally employed for defect sizing. The 6 dB drop, involves identifying the peak defect response and establishing defect boundaries based on a 50% energy dissipation threshold; giving it robust physics backed explainability [23], [24]. To account for varying industrial requirements, alternative amplitude drop thresholds (such as 12 dB or 20 dB) are also used [23], [25]. However, these lack the physical explainability of the 6 dB drop and typically require calibration through experimentation. The 6 dB drop is often extended and applied to amplitude C-scans for thresholding defect areas [26], [27], [28]. Whilst the 6 dB drop method is widely established it does have limitations [24], [26]. To tackle this challenge, previous research efforts have explored using Deep Learning (DL) approaches for defect sizing, employing fully supervised training of a 3D U-Net for volumetric defect segmentation [29]. This method, when compared to conventional amplitude drop sizing techniques, showcased notable improvements in defect sizing accuracy whilst demonstrating agreement with in-plane and through-thickness defect localisation.

While the 6 dB drop method remains a cornerstone in defect sizing, advancements in DL techniques offer promising avenues for enhancing defect localization and sizing accuracy. By leveraging the capabilities of fully supervised volumetric methods, researchers have made significant strides in addressing the complexities associated with defect characterization and quantification. Analysing volumetric data offers a wealth of information for defect characterization that surpasses what can be achieved through individual image analysis alone. Additionally, this approach proves advantageous in reducing manual preprocessing tasks, such as gating out structural responses, which are often labour-intensive. Moreover, volumetric segmentation masks of components open avenues for various downstream tasks by producing digital twins [30], such as Finite Element Analysis (FEA), and can significantly alleviate the burden of report generation.

However, these methods face limitations due to the necessity of large amounts of labelled training data, a common requirement for any fully supervised training approach. In many NDE applications, obtaining accurately annotated labelled datasets of real defects is challenging and often not possible, and is one of the main barriers for applying DL to NDE [21]. The limited availability of labelled data, alongside industry concerns about interpretability and the absence of relevant standards, presents significant challenges to the effective development and adoption of DL techniques in NDE. To address this scarcity issue, previous work utilised synthetic training data generated from simulations [31], [32], [33]. By controlling simulation parameters, accurate labelling of segmentation ground truth could be achieved automatically. However, accurately simulating the full distribution of defects and their variations is computationally demanding, and it is challenging to ensure fidelity between the simulated and experimental domains [31]. As a result, if defect segmentation could be performed without the reliance on large positively labelled datasets, it would offer significant advantages.

Self-Supervised Learning (SSL) [34], [35] is a method for training DL models in a supervised manner without the need for labelled training data. Generally, labels are generated through auxiliary tasks or by leveraging inherent structures in the data itself, enabling the model to learn meaningful representations without explicit human annotation. By training in a supervised manner, models are often able to learn more detailed feature representations than unsupervised approaches. SSL introduces the ability to leverage large amounts of unlabelled data, reducing the need for costly and time-consuming annotation. Its versatility has meant that it has been applied broadly from computer vision to natural language processing and has demonstrated impressive performance in many notable DL tasks, such as with large language models [36], [37] and large vision models [38], [39]. In general, SSL works by formulating pretext tasks that require the model to predict certain aspects of the input data based solely on the input itself. These pretext tasks are designed to be easily computable from the raw data without the need for external annotations. By solving these pretext tasks, the model learns to extract meaningful features and representations from the data, which can then be transferred to downstream tasks. For example, in natural language processing, the model may be tasked with predicting missing words in a sentence (e.g., masked language modelling [40]) or predicting the next word in a sequence (e.g., language modelling [36], [37]). Similarly, in computer vision, SSL tasks may involve predicting the rotation, colorization, or spatial arrangement of patches within an image [41], [42], [43], [44]. Another prevalent strategy in SSL is contrastive learning, where the model learns to differentiate between positive and negative pairs of data samples. By maximizing the similarity between positive pairs (e.g., different augmentations of the same image) while minimizing the similarity between negative pairs (e.g., augmentations of different images), such as Siamese networks [45].

This paper presents a novel approach aimed at revolutionizing the detection, localization, and segmentation of defects within ultrasonically inspected volumes. The approach leverages the capabilities of SSL coupled with a 1D (1-Dimensional) Convolutional Neural Network (CNN) to achieve 3D segmentation of defects from volumetric ultrasonic testing data of composite components. Pretext learning is employed to predict distributions of amplitudes from ultrasonic sequences. During the inference stage, the pre-trained 1D CNN is deployed to flag any regions within the component that exhibit anomalous behaviour. This process capitalises on the insights gleaned from the pretext learning task, where the model familiarises itself with clean samples. Unlike traditional approaches, which often necessitate extensive positive training examples, this methodology operates on the principle of anomaly detection rather than classification into specific defect classes. By reframing the problem as one of anomaly detection, rather than attempting to categorize defects into predefined classes, the significant challenge of acquiring an extensive set of positive training examples is circumvented. Moreover, this approach offers the advantage of being defect-agnostic, thereby mitigating concerns regarding the generalisability of the model to novel defects. This characteristic alleviates the need for meticulous fine-tuning and ensures that the model remains robust across various defect types, eliminating the burden of adapting the system for each new defect encountered. Overall, by adopting this innovative methodology, critical limitations in the application of DL to NDE are addressed, paving the way for more efficient and robust defect detection in ultrasonic inspection processes whilst giving information on not just detection but also complete volumetric localisation and segmentation which has only been previously achieved with fully supervised training [29].

Section 2 of this paper outlines the data used and any pre-processing requirements. In Section 3, the pretext learning is detailed. The results and discussion are presented in Section 4.s This, for the first time, introduces an SSL method for volumetric defect segmentation of ultrasonic testing data, offering several advantages:

- No labelled or defective training data is required.
- Minimal preprocessing (removal of Time-Compensated Gain (TCG), gating, and front-wall peak-alignment).
- Geometric features are retained.
- Enhanced generalizability and geometrical independency for defect detection is achieved by reframing the problem as anomaly detection.

## 2. Data

CFRP samples of varying thicknesses, supplied by Spirit AeroSystems, were employed in this study. For pretext learning, samples verified as defect-free through ultrasonic inspection, analysed by an NDE operator, were selected. These samples were segregated into distinct datasets for training, validation, and testing purposes. During the inference phase, defective samples were used for testing. The initial set of defective samples featured Flat-Bottom Holes (FBH). The first sample contained 15 flat-bottom holes with diameters of 3.0, 6.0, and 9.0 mm. These holes were drilled to depths of 1.5, 3.0, 4.5, 6.0, and 7.5 mm from the inspection surface. The second sample featured 25 flat-bottom holes, all of which were drilled to the same depths as those in the first sample, with additional defect diameters of 4.0 and 7.0 mm. This allowed for a range of defect sizes and positions to be tested. A final stepped sample with square 6 mm wide Polytetrafluoroethylene (PTFE) inserts presented a more challenging geometry and defect responses, offering a representation closer to naturally occurring defects. A summary of the samples and dataset characteristics is presented in Table 1.

Data acquisition was conducted using an automated robotic phased array ultrasonic system centred around a 64 element, 5MHz Olympus linear phased array roller probe, with an element pitch of 0.8 mm [46] and 100 MHz sample rate. The components were scanned using a raster pattern with a sub-aperture of 4 elements, to ensure sufficient acoustic energy, giving 61 individual beams per pass. Further details regarding the experimental setup and composite samples can be found in [32]. The segmentation methodology employed in this study was intentionally designed to be versatile, thereby maximizing its applicability across different scenarios. Consequently, only minimal generic data preprocessing was applied. The data preprocessing involved enveloping the signal using the Hilbert Transform. Taking the envelope of the signal is a valuable tool for extracting the instantaneous response within a time series, and is commonly employed in generating C-scan images from ultrasonic A-scans [47]. Notably, techniques such as TCG, peak-alignment, and gating out of geometric features were deliberately omitted.

*Table 1: Summary of samples used.*

| | Sample | Thickness (mm) | Dataset Size [Probe \| Time \| Frames] | Details |
|---|---|---|---|---|
| *Pretext Learning* | Clean 1 | 2.75 | 122\|350\|260 | Training |
| | Clean 2 | 4.25 | 122\|450\|260 | |
| | Clean 3 | 4.25 | 122\|450\|260 | |
| | Clean 4 | 6.00 | 122\|600\|260 | |
| | Clean 5 | 6.00 | 122\|600\|260 | Validation |
| | Clean 6 | 8.60 | 122\|700\|260 | Test |
| *Inference* | Defective 1 | 8.60 | 183\|700\|260 | 15 FBH |
| | Defective 2 | 8.60 | 305\|700\|230 | 25 FBH |
| | Defective 3 | 7.50,9.60,11.80 | 488\|1050\|112 | 15 Inserts |

## 3. Pretext Learning

Typically, ultrasonic signals are considered as time-traces, which is often appropriate. However, when inspecting large components C-scan images often provide the most impactful and clear information about the structure of the component. C-scans compress information in the time-domain either through amplitude imaging or amplitude indexing (Time-of-Flight). This loss in temporal information is done to maximise comparative spatial information, which can often be the most helpful when evaluating components of known geometry, particularly for composite components where structural noise can show significant variability in the time-trace. Operators will often produce C-scan images at different thickness gates to compare different through-thickness slices. Intuitively therefore it can be deduced that considering comparative spatial information for given time-windows is as, if not more, important than direct temporal comparisons along a single time trace. Ideally spatial comparisons would be conducted at every depth but for manual interpretation this is intractable due to the substantial amount of data produced.

Mechanized linear phased array scanning is commonly employed in the inspection of large-scale industrial components [15], [16]. Arrays operate within acceptable tolerances of element sensitivity. Whilst compensation can be done to account for inter-element variations within arrays through calibration, some level of variation between elements is likely to remain during scanning. Considering this and the importance of spatial comparison discussed earlier, it follows that it is appropriate to analyse a singular series through a component (parallel to the scan direction) at a particular time step for a given beam (Figure 2), and is the basis for this work. This series is hence referred to as a scan sequence. This approach minimizes variations from array elements by concentrating on the same transmit/receive element or sub-aperture, while facilitating spatial comparison at a given depth. Furthermore, this methodology can be easily conceptualized as a self-supervised learning task. By treating it as a 1D task, there was not only a reduction in model size and complexity but also a substantial increase in training data abundance compared to using 2D or 3D training sets from the same sample availability.

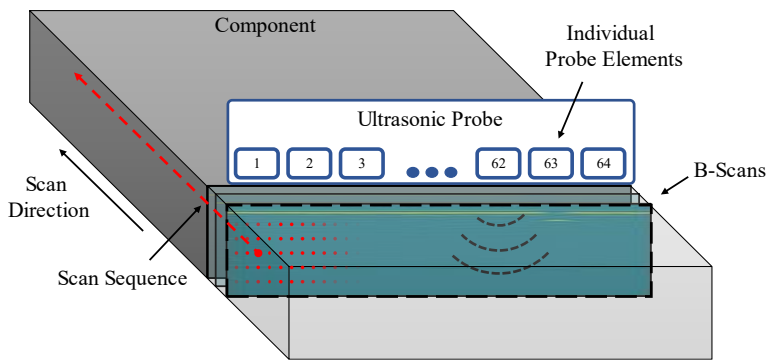

*Figure 2: Diagram illustrating an example of a through-component scan sequence, as examined for the pretext learning task.*

For the pretext training task in this study, inspiration is drawn from language modelling (where a model tries to predict the next word in a sequence), aiming to predict the next value in the scan sequence for a clean sample. While alternative methods such as contrastive learning or generative learning could be utilised, the task of predicting the next value lends itself well to in-process inference and, more broadly, to the downstream inference task of volumetric segmentation.

A Probabilistic Neural Network (PNN) was employed for sequential prediction to allow the model to account for differences in response variability. In this approach, the model attempts to predict the distribution that corresponds to the next value in the sequence based on prior information, as depicted in Figure 3. This enables the model to account for areas of prior variability or lack of variability by widening or tightening the distribution. Thresholds can then be set as confidence intervals against these distributions, automatically accommodating scan sequence variability. A scan sequence length of 64 values was employed for prediction, hence known as an input sequence, aiming to strike a balance between having a sufficiently large receptive field to learn about the distribution and patterns of amplitude responses within the component, while also ensuring that the sequence length is not excessively long, which could limit inference to large components and minimize access to training data.

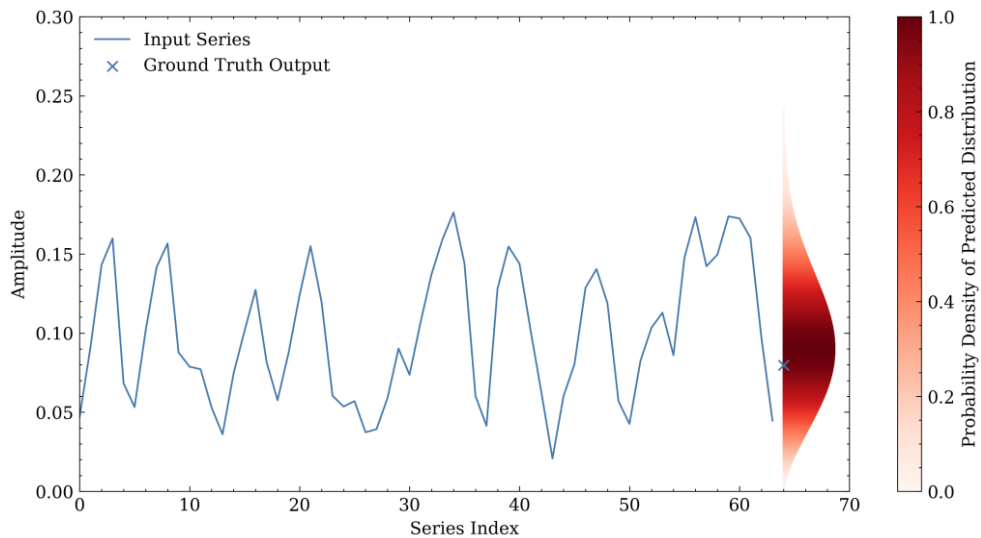

Figure 3: Example of predicted distribution from an input sequence.

Various models for series forecasting exist, ranging from DL methods such as Recurrent Neural Networks (RNNs) and Transformers to probabilistic approaches like Gaussian Processes [48], which can be adapted for probabilistic prediction based on specific requirements. In this study, a 1D multi-head CNN, inspired by InceptionTime [49], was adopted for the architecture, with short series (<128) shallow networks with a lower number of shorter filter lengths being deemed sufficient. This has enabled the 1D CNN to be a lightweight model, with 486,242 trainable parameters occupying just 1.94 MB of memory.

The model probabilistic approach was characterised by predicting the scale and concentration parameters of a two-parameter Weibull distribution as outputs. The choice of a two-parameter Weibull distribution enabled the modelling of various distribution shapes for continuous positive values (as consistent with enveloped amplitude values), accommodating non-symmetric distributions through both left and right skewed data. While this model and method proved effective for our application, the exploration of alternative probabilistic regression methods and architectures has been left for future work. The model architecture is depicted in Figure 4, where each convolutional block consists of a convolutional layer with the specified kernel size and stride of 1, followed by a convolutional down-sampling layer with a stride and kernel size of 2. A LeakyReLU activation function was employed between each convolutional and fully connected layer.

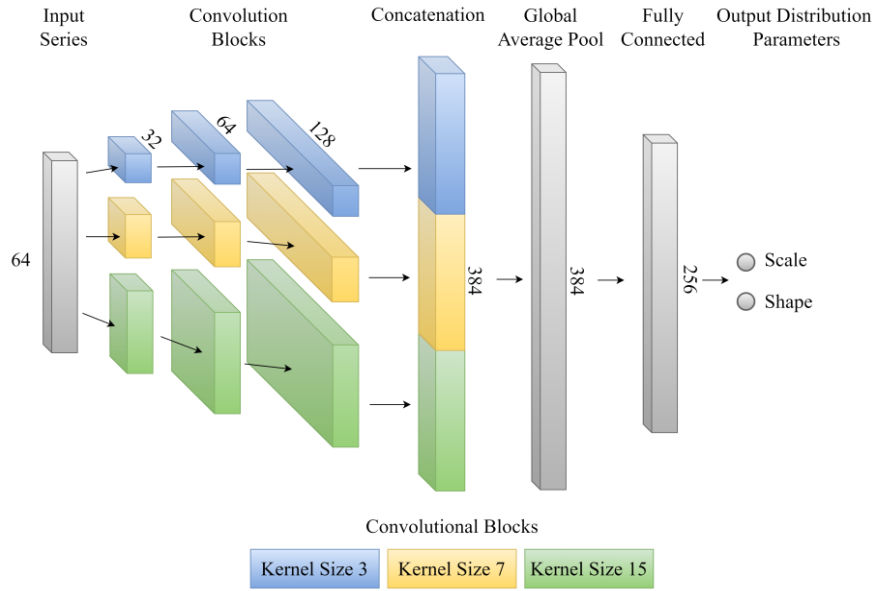

Figure 4: Probabilistic CNN architecture.

During pretext learning, clean samples 1-4 were used for training with a batch size of 65536, whilst clean samples 5 and 6 were kept as holdouts for the validation and test sets respectively (Table 1). Adam optimiser [50], with a learning rate of $1 \times 10^{-6}$ was used to minimise the Negative Log-Likelihood (NLL) loss for the Weibull distribution, as given by equation (1), where $f(a, b \mid x_i)$ is the Weibull probability density function parameterised by Scale ($a$) and Shape ($b$). To minimise overfitting, a patience of 3 epochs was used when evaluating the validation set to determine early stopping.

$$Weibull\ NLL\ Loss = -\log \prod_{i=1}^{n} f(a, b \mid x_i) = -\sum_{i=0}^{n} \log f(a, b \mid x_i) \qquad (1)$$

During training the data was down sampled in the time domain by every 5 samples (approximately 30 µm in depth within the CFRP samples). This was done as sequences next to each other in the time domain are almost identical; offering limited additional information to be learnt and an increased computational cost during training.

During training, a hyperparameter which arises for this problem is the stride for sampling data during training. Consider a single full-length scan sequence: the rate at which this full length is sampled for new training samples is determined by the stride of the window applied to each input sequence, as demonstrated in Figure 5. In DL, it is generally advantageous to maximize the amount of training data available. Using a stride of 1 achieves this by providing maximum available training data. However, whilst each sample is different by a single point, there exists significant overlap between neighbouring input sequences, exposing the model to very similar data during training. In such cases, the model may not gain additional information from the additional samples and could become prone to overfitting the training data.

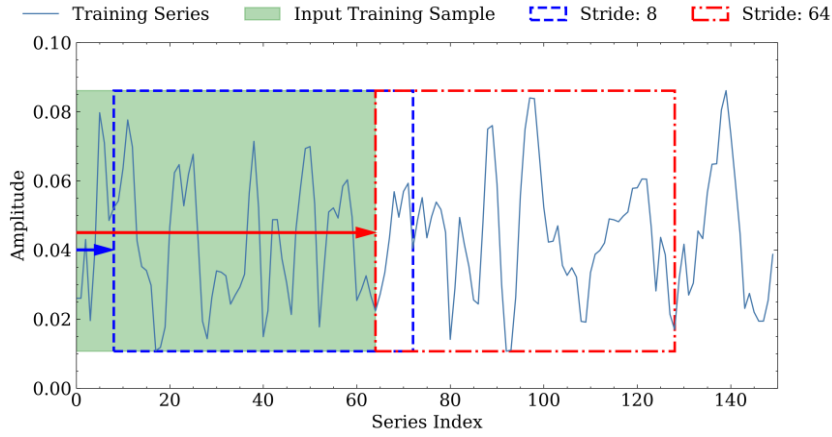

*Figure 5: Demonstration of the impact of stride (8 and 64) when sampling the training data.*

Alternatively, a stride length of 64 can be employed, where each training example represents a distinct input sequence without information sharing due to overlap with other training examples. While this reduces overlap, it also significantly reduces the available training data. Picking an appropriate stride for sampling training input sequences is therefore a trade-off between training set size and overlap in the training data, potentially leading to overfitting. To explore this relationship, the following stride values were tested: 1, 2, 4, 8, 16, 32, 64, 128, 256 which corresponded to dataset sizes of 74.12 M, 37.06 M, 18.53 M, 9.46 M, 4.92 M, 2.65 M, 1.51 M, 0.76 M, 0.38 M respectively. It is worth highlighting that an advantage of using 1D sequential data in this way is the production of very large training datasets compared to typical image-based ML research in NDE. For this test, the stride of the validation set was fixed at 64, and the testing stride was set to 1 to maximize the test set size (which also matches the requirement during inference). The model was trained three times for each stride, and the average results were taken. The outcomes are presented in Figure 6.

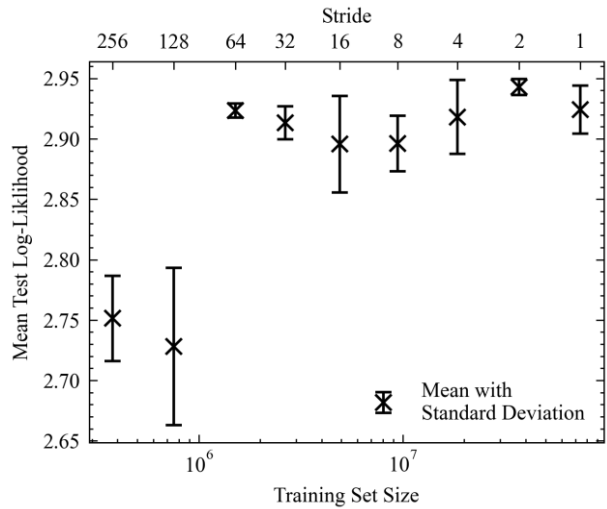

*Figure 6: Test set mean Log-Likelihood for varying sampling strides.*

As depicted Figure 6, a stepped improvement is observed when the training datasets exceed 1 million samples (stride < 128). However, beyond this threshold, the model appears to exhibit diminishing returns from additional training data. Similar trends to this have been documented in the existing literature [51]. Perhaps surprisingly, there was not a consistent increase in performance for a reduction in stride. Whilst this could indicate model saturation, there could also be a detailed relationship between training data and overfitting at play. Notably, strides of 64 and 2 exhibited lower deviations in results, coupled with good performance, suggesting the emergence of stable solutions. It's important to note that different pretext learning models and training datasets may exhibit varying relationships with stride length. Nonetheless, this analysis underscores the significance of tuning this parameter and provides insight into the required training dataset sizes. Future work hopes to explore this more and assess its impact on inference performance.

# 4. Inference
## 4.1. Methodology

By training on defect free data, the pretext learning phase yields a model which predicts the expected distributions of the next defect free value for a given time index and reception element. By leveraging this pre-learnt information, the 1D model is utilized by applying it to each preceding 64 long input sequence of volumetric data to forecast the distribution for each point in the subsequent B-scan. This enables anticipation of the distribution of data points across the entire next B-scan under the assumption of a defect-free component.

During inference each volumetric frame (B-scan) was processed in a sequential manner, comparing the predicted clean B-scan to the measured experimental result at every point in time for each receptive element or sub-aperture of elements. Since the model was trained only on clean data to predict the next clean value in the sequence, if there was agreement between the values in the predicted frame and the measured experimental frame they were marked as defect free. The inference window was then moved to the next frame in the data and the process continued. This acts to update the predictive model with the most current information and keeps good alignment with variations seen in specific samples/scans.

However, if the next B-scan contains defective voxels this will result in an increased amplitude response around the defect which is outside the distribution of clean sequences seen in the prior learning. As a result, when comparing the predicted (clean) frame to the measured (defective) frame there will be poor agreement locally around the defect response. These voxels can therefore be marked as defective within the volume; locally segmenting the defect. For the sequences used for the subsequent frame, prior predictions not marked as defective are treated as normal with the input sequence window sliding along from the experimental scan sequence. However, for defective voxels the mean is taken from the probabilistic output as the most likely value for a clean response and is used to update the following sequence. This ensures that the model is always predicting clean responses based on prior clean sequences and defective responses do not impact future predictions, as consistent with the pretext learning. This process is outlined in Figure 7.

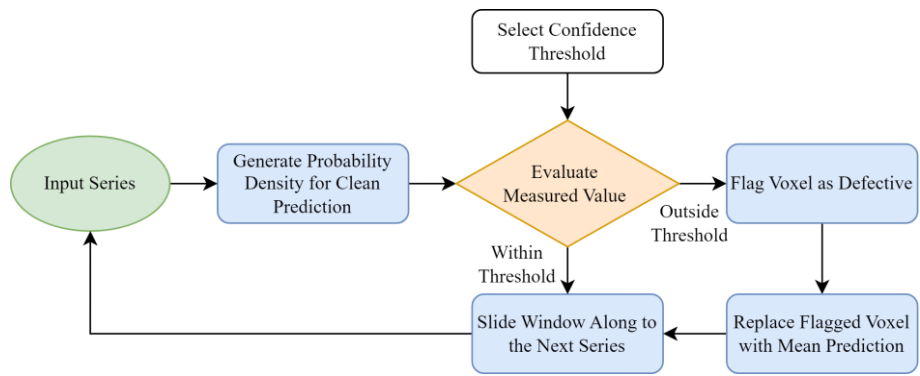

*Figure 7: Flowchart of sequence prediction for inference.*

Given the probabilistic output, a threshold must be set to evaluate the model's clean prediction against the measured value to determine if a voxel is defective. This threshold can be set as a confidence threshold on the predicted clean output expectation, making it adaptive to variations in input sequence amplitude responses. There are different ways to determine an appropriate confidence threshold such as experimental calibration. An alternative is to base it off an allowable false-call rate. Since this detection analysis is for the entire volume and due to the resolution difference in the spatial and time domain, the number of voxels is far larger (700 times for test samples 1 and 2) than for image level analysis. For the same expected number of absolute false calls for image analysis the false call rate for the volume would therefore need to be far lower (due to the increased number of voxels). To account for this a confidence threshold is chosen based on a much lower allowable false-call rate than would be expected for image analysis. It is important to note that this false-call rate is done on a per voxel basis and not per defect basis. In this study results are presented for false-call rates ranging from 1% to 0.00001%. Even at the aggressive lower false-call rate of 0.00001, sample 2 which has 5.6 M voxels would still be expected to produce approximately 6 false-calls. Whilst for these samples this is a low number, for larger parts this would scale cubically for any increase in scan length.

For volumetric inspection, the inference process can be completed just following the forward scan (as most applicable to in-process inspection), herein denoted as the forward sweep, or post scan the analysis can be completed in reverse – simulating the scan from the other direction, herein denoted as the backwards sweep. Taking the logical AND of both passes as the final segmented volume acts to increase confidence in defective predictions by having them detected twice. This helps to further remove false positive indications and cleans up the segmented response as seen in Figure 9.

To further mitigate the impact of false-calls, a minimum area threshold was applied to detect only defects above a critical size and limit low-area voxel clusters resulting from the probabilistic nature of predictions. For CFRP applications, defects generally align parallel to the ply orientation. It is this area which is typically used to assess critical defect size. Area thresholding was calculated using equation (2), where *Area Opening* refers to the *skimage.morphology.area_opening* [52] which, removes all connected components smaller than the *Filter* for the volume. The connectivity determines what neighbours are considered connected components, for this application all 8 neighbouring voxels are considered for a spatial plane through the volume. For other applications and materials, where the defects are not primarily in-plane, 3D connected components may be more appropriate.

$$Thresholded\ Volume = Area\ Opening(Volume[:, depth, :], Filter, Connectivity) \quad (2)$$

The minimum defect size threshold can be tailored based on the specific application requirements. For testing purposes, where the minimum defect size was 3.0 mm in diameter, the area threshold was adjusted to exclude any indications smaller than this. Area segmentation is completed as the final processing stage. The complete workflow for volumetric instance segmentation is illustrated in Figure 8. Details on the impact of each processing step for different thresholds can be found the in the following Results section (Figure 11) and Table 4 of the Appendix.

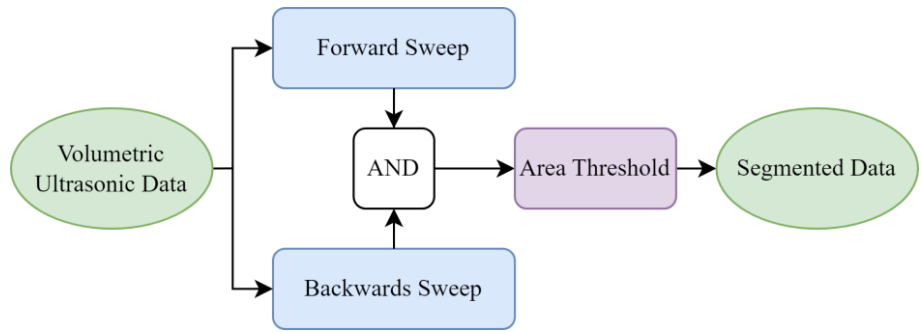

*Figure 8: Flowchart of the methodology overview for complete volumetric segmentation.*

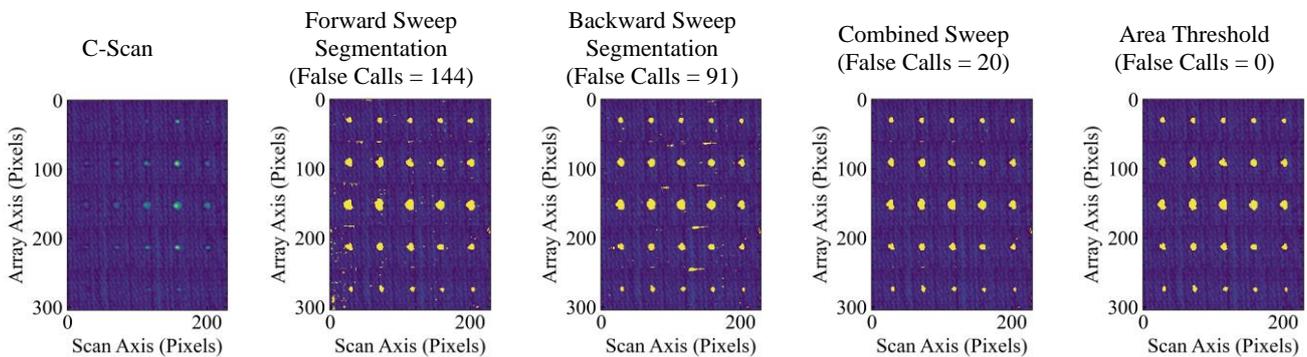

*Figure 9: Demonstration of the impacts of different post-processing steps for defect sample 2 and threshold of 0.9999999.*

To process the first frames where there are not 64 prior frames for prediction, padding is used to populate the remaining missing information for each sequence. Whilst zero padding, constant edge padding or reflect padding can all work as the combined sweeps clear up any errors as a result of not having enough prior information for predictions. They can produce artefacts for a single sweep at the edge of the scan due to a lack of prior information for the model to give an accurate prediction. Reflect and edge padding give the best results, reducing the reliance on sweeps to clean up edge artefacts. Figure 10 shows an example of edge artefacts from using zero padding compared to edge padding.

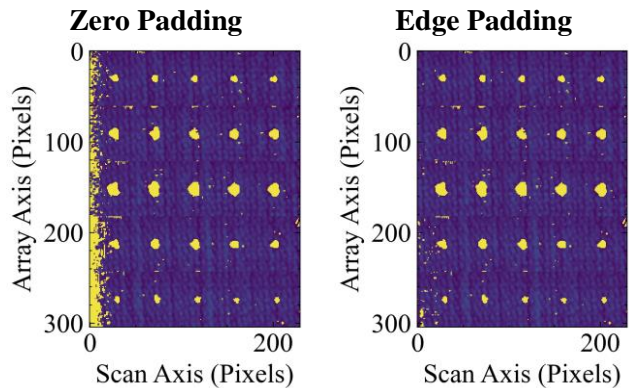

*Figure 10: Example of edge artefacts for a single sweep when using zero padding compared to edge padding (sample: defect 2, threshold: 0.9999999).*

Different defects exhibit varying characteristics. This method reframes the problem of positive defect identification as one of anomaly prediction, which is beneficial for generalisability to a wide range of defects. However, as the model has not learnt specific positive features of defect responses, any change in amplitude response, whether from a defect or geometric feature, can lead to an anomalous prediction. Real components often feature geometric elements such as step changes in thickness. While inference can still be conducted in such cases, analysis must be conducted in parallel with any geometric alterations so that they are present in the sequence used for prediction. Therefore, if there is a part with very complex geometry it may have to be sectioned for inference. To evaluate the method's performance under these conditions and to assess its effectiveness with different defect types, the approach was applied to a stepped sample with PTFE inserts (defect sample 3). These defects are inserted pre-cure and are more representative of naturally occurring defects compared to FBHs.

For inference the model used was the best performing during pretext learning. Inference is a sequential process, but the computational cost can be easily minimised by batching the predictions for a complete frame; due to the lightweight 1D CNN. To reduce the computational cost during inference, the time domain was down sampled by a factor of 10 for testing. Previous work demonstrated that it is still possible to get very high levels of depth-wise resolution with this sampling [29]. Inference for a single frame took approximately 0.05s when batched. To complete the full processing pipeline of forwards and backwards sweeps and area thresholding took less than 35s for each sample. Testing was conducted on a workstation equipped with a NVIDIA GeForce RTX 3090 running the Pytorch [53] framework. The following section presents results for defect detection, sizing, localisation, and visualisations for qualitative examples of volumetric segmentation.

### 4.2. Results

*Detection*

Detection performance of the model was evaluated for thresholds ranging from 0.99 to 0.9999999, with results reported for each processing stage. Evaluation was conducted on amplitude C-scans of the segmented volumes to avoid the impact of repeat echoes stemming from defect indications. The ground truth defect mask was generated using manual identification of defects, with the 6 dB drop used for locating the centroid of each defect. The true defect sizes were then used around the centroids to mark out the defect areas. A defect is considered detected if the predicted mask overlaps with the ground truth mask. If there is no overlap the prediction is considered a false-positive. Across all thresholds and processing stages, all defects were successfully identified, with no missed detections; however, a notable decline in accuracy is observed as a result of false positive indication. Whilst the absolute number of false positives can be large, they are often very small (as demonstrated in Figure 9) and can therefore be effectively removed with area thresholding. The results for detection accuracy are presented in Figure 11 (with a full breakdown available in Table 4 in the appendix). It is feasible to completely minimize false positives, thus achieving a detection accuracy rate of 100%, using the processing steps suggested in Figure 8 and a threshold of 0.9999999.

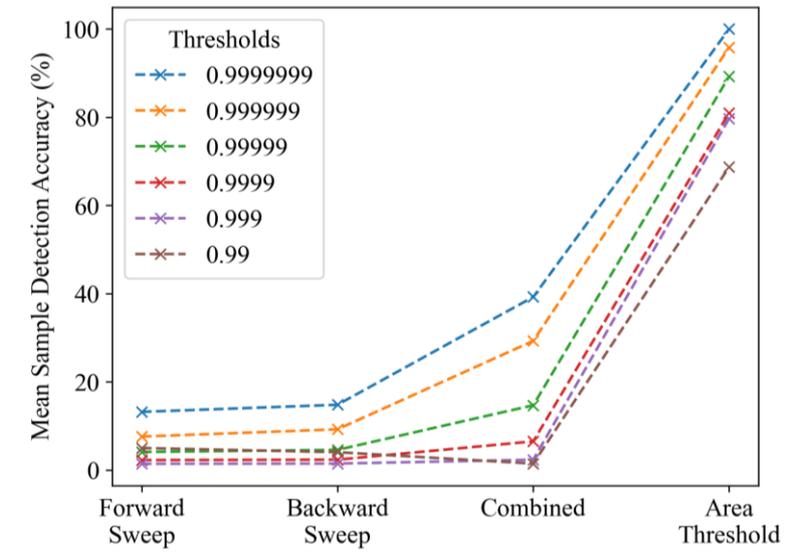

*Figure 11: Defect detection accuracy for each threshold and processing step.*

*Defect Sizing*

The impact of defect sizing has been assessed across the same range thresholds as used for detection, with the results summarised in Table 1. While the deviation of absolute error remains relatively stable, there exists a strong negative correlation between increasing the confidence threshold and a reduction in MAE for defect sizing. Notably, at a threshold of 0.9999999, the MAE of 1.41 mm, aligns closely with the findings of a previous study which reported errors in sizing inaccuracy for the 6dB drop method of 1.35 mm [29]. Moreover, this performance surpasses that of the fully supervised 3D U-Net method previously presented, prior to adjustments for differences between synthetic and experimental domains.

The method has been shown to only oversize defects. This stems from the underlying mechanism of detecting anomalous voxels, which does not always correspond perfectly in a one-to-one manner to actual defect size; due to impacts of ultrasonic imaging such as beam spread etc. The impact of this is visually demonstrated through an example B-scan in Figure 12. Consistency in oversizing defects yields two main advantages:

**Safety Considerations**: Oversizing defects, while potentially sacrificing some precision, inherently reduces the risk of overlooking critical flaws. In safety-critical industries such as aerospace or structural engineering, the consequences of under sizing defects can be severe, leading to structural failures or operational hazards. By consistently erring on the side of caution and oversizing defects, the method provides a safety buffer, ensuring that potential weaknesses are identified and addressed before they escalate into safety incidents.

**Calibration Potential**: The method's consistent error pattern facilitates calibration procedures, akin to those commonly employed in traditional amplitude drop sizing methods. Calibration involves establishing a correlation between measured responses and actual defect sizes, thereby refining the method's accuracy and reliability. The method's predictable oversizing behaviour lends itself well to calibration processes, allowing for adjustments that compensate for systematic errors. For instance, for a threshold of 0.9999999, when utilizing defect sample 1 as a calibration reference, the Mean Absolute Error (MAE) for Defect Samples 2 and 3 decreases to 0.58 mm, a 57% reduction in MAE.

Table 2: MAE for different thresholds.

| Sample | Defect Width (mm) | MAE for given threshold (mm) | | | | | |
|---|---|---|---|---|---|---|---|
| | | 0.99 | 0.999 | 0.9999 | 0.99999 | 0.999999 | 0.9999999 |
| Defective 1 | 9 | 2.75 | 2.25 | 1.41 | 1.27 | 1.13 | 1.02 |
| | 6 | 3.39 | 2.55 | 2.05 | 1.67 | 1.44 | 1.27 |
| | 3 | 3.78 | 2.91 | 2.60 | 2.21 | 1.83 | 1.75 |
| Defective 2 | 9 | 3.14 | 2.38 | 1.86 | 1.49 | 1.20 | 0.93 |
| | 7 | 3.26 | 2.61 | 2.19 | 1.83 | 1.50 | 1.23 |
| | 6 | 3.18 | 2.62 | 2.10 | 1.74 | 1.37 | 1.23 |
| | 4 | 3.67 | 2.97 | 2.45 | 2.17 | 1.93 | 1.79 |
| | 3 | 3.93 | 3.39 | 2.94 | 2.58 | 2.32 | 2.06 |
| Defective 3 | 6 | 3.06 | 2.32 | 1.73 | 1.46 | 1.29 | 1.21 |
| **Mean** | | **3.37** | **2.71** | **2.20** | **1.87** | **1.59** | **1.41** |
| **Standard Deviation** | | 0.66 | 0.61 | 0.69 | 0.67 | 0.66 | 0.68 |

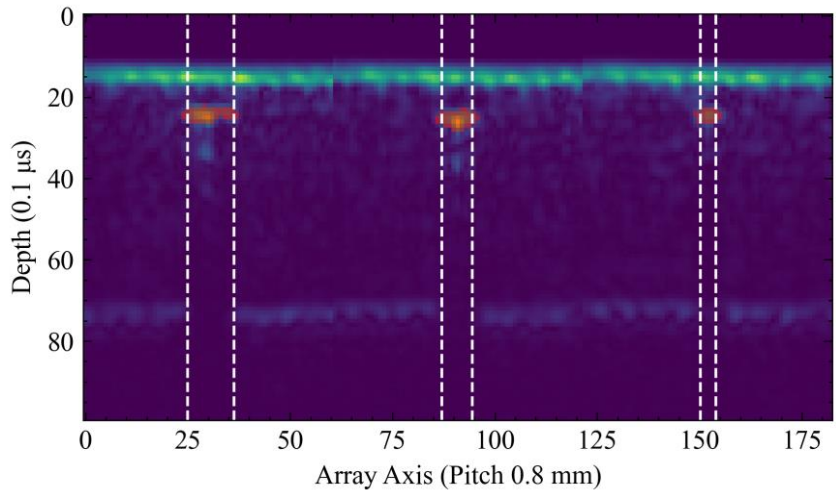

Figure 12: Example B-scan across multiple raster passes showing the voxels highlighted as defective and the corresponding true defect size in white. This is shown for defect sample 1 with a threshold of 0.9999999.

*Localisation*

For linear scanning of composite materials, achieving accurate volumetric localisation necessitates an understanding of both in-plane and depth-wise localisations. For depth-wise localisation, the evaluation focuses on FBHs, while excluding inserts. Inserts, positioned between layers pre-cure, pose a challenge in obtaining true ground truth depth measurements post-curing, unlike FBHs which can be accurately measured post cure to obtain a ground truth depth measurement. Through-thickness measurements are derived from the mean thickness through the centroid of the segmented defect.

In contrast, acquiring a ground truth measurement for in-plane localisation presents its own challenges. Cumulative positional errors inherent in the experimental setup render the attainment of accurate ground truth measurements (<1.0 mm) infeasible. Consequently, this research adopts the 6 dB drop criterion as a reference standard for validating the agreement between segmented in-plane localisation and the 6 dB drop criterion. By doing this the in-plane localisation can be compared against an industry standard method as a validation benchmark. To assess in-plane localization, the Euclidean distance between the centroid of the 2D projected segmentation mask and the 6 dB drop masks is computed.

The integration of both through-thickness and in-plane localisation provides a comprehensive understanding of the method's ability to locate defects within the volume. Table 3 presents the mean and standard deviation of results. The MAE for depth localisation surpassed that of in-plane localisation, with comparable standard deviations for both. Although this method showed improvement for in-plane localisation compared to previous fully supervised volumetric segmentation methods, it fell short of achieving the high accuracy levels seen in through-thickness depth localisation. This discrepancy might be partly attributed to the additional pre-processing steps used for the fully supervised method, such as peak alignment. Enhancing through-thickness accuracy could potentially be achieved by increasing the temporal sampling rate during inference, although for most applications, the current level of accuracy is likely sufficient. Both metrics demonstrate considerable localisation performance, with MAEs well below 0.5 mm. This level of precision is likely more than sufficient for typical industrial rework scenarios, as it aligns with the accuracy required for precise tooling operations. Additionally, the in-plane MAE is far below the element pitch, which is the limiting factor for in-plane imaging resolution. All in all, the model performs well in volumetric defect localisation.

*Table 3: Localisation results.*

| Sample | Defect Width (mm) | Depth (mm) | | In-Plane Distance (mm) | |
|---|---|---|---|---|---|
| | | MAE | Standard Deviation | MAE | Standard Deviation |
| **Defective 1** | 9 | 0.34 | 0.29 | 0.41 | 0.11 |
| | 6 | 0.10 | 0.05 | 0.50 | 0.28 |
| | 3 | 0.16 | 0.05 | 0.26 | 0.16 |
| **Defective 2** | 9 | 0.33 | 0.15 | 0.28 | 0.15 |
| | 7 | 0.30 | 0.08 | 0.32 | 0.10 |
| | 6 | 0.23 | 0.17 | 0.44 | 0.14 |
| | 4 | 0.23 | 0.15 | 0.31 | 0.08 |
| | 3 | 0.36 | 0.06 | 0.40 | 0.16 |
| **Defective 3** | 6 | - | - | 0.27 | 0.18 |
| **Total** | | **0.26** | 0.17 | **0.37** | 0.18 |

*Visualisations*

Volumetric segmentation offers a key advantage over image-based segmentation by providing comprehensive localisation information about defects. This enables more thorough visualisations and reconstructions, which can be particularly advantageous for constructing digital twins of components for testing or reporting purposes. Figure 13 presents visual examples of the volumetric segmentation results, showcasing the model's performance across various defects and stepped samples. Despite the absence of TCG and significant variations in defect response levels at different thicknesses due to attenuation, the method manages to deliver relatively consistent segmentation masks at different depths for defects of the same size by considering local temporal samples during inference. This resilience to variations in thickness and attenuation levels enhances the reliability and applicability of the segmentation method across a range of scenarios.

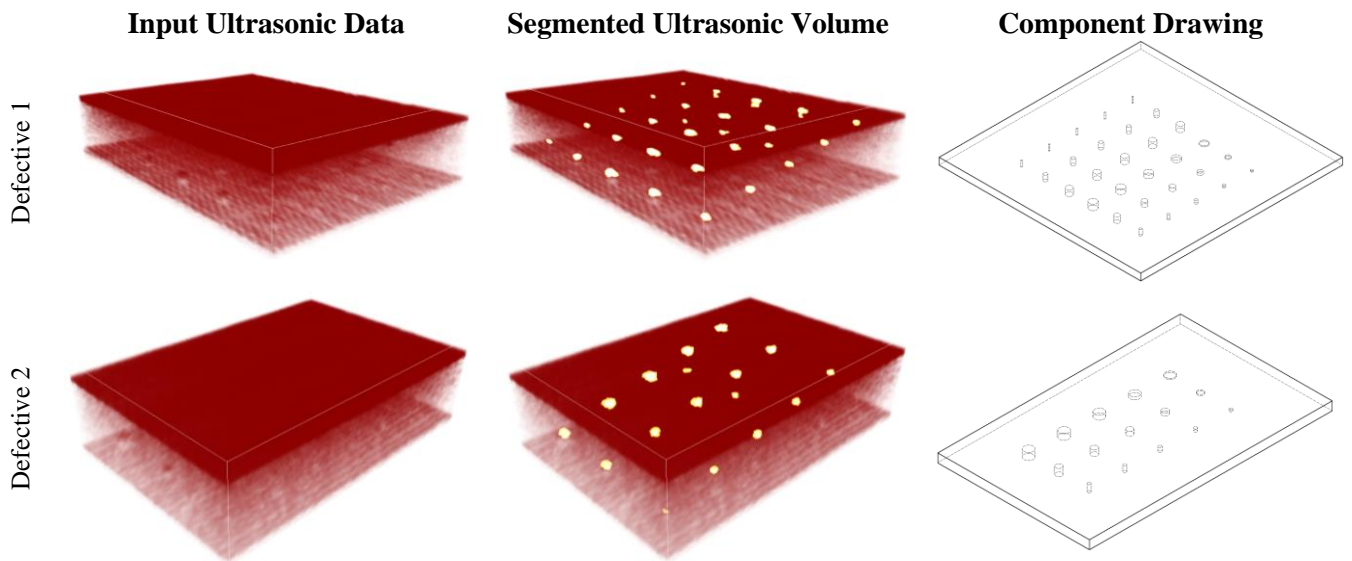

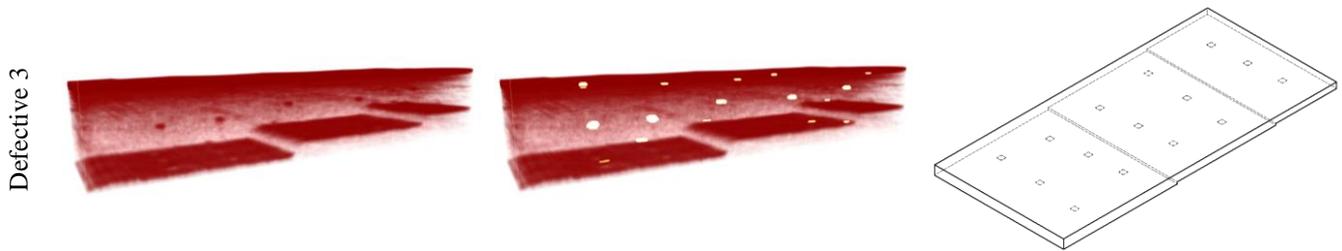

*Figure 13: Visualisations of volumetric ultrasonic responses, their corresponding overlayed segmentations, and component drawing.*

## 5. Conclusion

This paper introduces a new approach for ultrasonic volumetric defect segmentation using SSL to address the need for labelled training data. The method has been demonstrated for CFRP composites over different samples, defects, and geometries.

Volumetric segmentation provides information on defect sizing and localisation and allows for advanced visualisations which can facilitate the creation of digital twins for reporting or testing. One of the biggest challenges when applying DL methods to NDE is the requirement for training data. This is compounded for volumetric segmentation tasks as volumetric training data is even less available and segmentation models typically require highly intensive labelling at a per voxel level.

This method employs a 1D CNN to learn the expected distributions for clean data along a given time index for a receptive element. This learned information is then used to generate a volumetric defect segmentation mask. Several advantages stem from this approach. Firstly, it circumvents the need for expensive and challenging-to-obtain labelled training data. By simplifying the pretext task to a 1D data sequence, the accessibility of training data significantly increases; with millions of training samples derived from just four samples. Since this training data is defect-free, the limitation of obtaining large quantities of defective components no longer applies. Moreover, the size of the model used was only 1.94 MB. This means that although batching during inference is preferred for time-saving purposes, it is possible to reduce batch sizes or eliminate them altogether to allow the model to run on machines with more limited hardware.

Additionally, by framing the problem as one of anomaly prediction rather than positive defect identification, the method becomes more generalisable and robust to various types of defects, as long as they exhibit an amplitude response that deviates from the expected. This was demonstrated through testing on more challenging to detect PTFE inserts which were more accurately sized than the FBH. Furthermore, by treating the task as one of probabilistic sequence prediction, the segmentation approach becomes more interpretable compared to using a 3D end-to-end segmentation model. This aids in demystifying some of the "black box" nature of DL predictions, which is a significant challenge in the NDE industry.

All defects were reliably detected across thresholds and processing steps. False positive indications were successfully eliminated for a 100% detection accuracy with the complete processing pipeline and a threshold of 0.9999999. The results showcased good in-plane and through-thickness localisation, exhibiting improvements over a previously utilised fully supervised model [29] for in-plane localisation, with MAEs of 0.37 and 0.26 mm, respectively. Although through-thickness localisation performed slightly worse than previously demonstrated supervised model, it remains suitable for most applications, and it still falls below the in-plane localisation error. MAEs for defect sizing were also presented, revealing a negative correlation between sizing error and threshold. For a threshold of 0.9999999, the MAE aligned with the 6 dB drop at 1.41 mm. Due to the method's nature of detecting anomalous voxels in the ultrasonic domain, which may not always correspond directly to sizes in the physical domain, achieving accurate sizing remains challenging. However, the consistent over-sizing of defects enabled consistent calibration, which lead to a 57% reduction in sizing error.

However, there are limitations to this method. Firstly, the method requires one consistent geometric axis to be applied to the data. Whilst its robustness was demonstrated on stepped samples, more complex shapes may prove challenging, and the method would likely require modifications to accommodate such complexities. Secondly, most DL models shift the computational load of understanding a task to training, which can be done offline, prior to inference. This typically results in rapid inference results – a key benefit over other interpretation methods.

However, the sequential nature of inference for this method means that there is an increased computational cost and time during interpretation. Although the full interpretation time was minimized to a maximum of 35 seconds using input sequence batching for each B-scan, for less powerful machines, this may not always be feasible. Additionally, for larger parts, the inference time will increase as a result of longer scans, however inference is still likely to be far quicker than human inspection in real-world applications. Future work aims to explore the impact of different models for pretext learning, different materials and scanning methodologies, and apply the method to in-process inspection.

## 6. Acknowledgment

This work was supported through Spirit AeroSystems/ Royal Academy of Engineering Research Chair for In-Process Non-Destructive Testing of Composites, RCSRF 1920/10/32.

## 7. Appendix

*Table 4: Detection accuracy across thresholds and processing steps for each sample.*

| Threshold | Sample | Detection Accuracy % (False positives) | | | |
|---|---|---|---|---|---|
| | | Forward Sweep | Backward Sweep | Combined Sweep | Area Threshold |
| 0.9999999 | 1 | 16.85 (74) | 12.40 (106) | 39.47 (23) | **100.00 (0)** |
| | 2 | 14.79 (144) | 21.55 (91) | 55.56 (20) | **100.00 (0)** |
| | 3 | 7.98 (173) | 10.49 (128) | 22.73 (51) | **100.00 (0)** |
| 0.999999 | 1 | 9.20 (148) | 7.54 (184) | 28.85 (37) | 93.75 (1) |
| | 2 | 8.28 (277) | 13.30 (163) | 42.37 (34) | 100.00 (0) |
| | 3 | 5.34 (266) | 6.91 (202) | 16.67 (75) | 93.75 (1) |
| 0.99999 | 1 | 4.66 (307) | 3.83 (377) | 12.83 (102) | 83.33 (3) |
| | 2 | 4.28 (559) | 5.94 (396) | 20.49 (97) | 96.15 (1) |
| | 3 | 3.42 (395) | 3.99 (337) | 10.79 (124) | 88.24 (2) |
| 0.9999 | 1 | 2.17 (677) | 2.02 (729) | 5.68 (249) | 71.43 (6) |
| | 2 | 2.33 (1046) | 2.76 (880) | 7.99 (288) | 92.59 (2) |
| | 3 | 2.33 (586) | 2.34 (584) | 5.88 (240) | 78.95 (4) |
| 0.999 | 1 | 1.31 (1129) | 1.36 (1085) | 1.89 (780) | 65.22 (8) |
| | 2 | 1.33 (1704) | 1.39 (1780) | 2.32 (1053) | 89.29 (3) |
| | 3 | 1.68 (759) | 1.71 (806) | 2.85 (512) | 84.62 (2) |
| 0.99 | 1 | 4.95 (288) | 4.79 (298) | 1.51 (980) | 50.00 (15) |
| | 2 | 5.73 (411) | 3.81 (632) | 1.30 (1896) | 96.15 (1) |
| | 3 | 4.34 | 3.60 | 1.62 | 60.00 |

| | | (331) | (402) | (913) | (10) |
|---|---|---|---|---|---|